\documentclass[draft]{agujournal2019}
\usepackage{url} 
\usepackage{lineno}
\usepackage[inline]{trackchanges} 
\usepackage{soul}
\journalname{Geophysical Research Letters}

\begin{document}

\title{Challenges in determining whether ENSO is a damped or a self-sustained oscillation}

\authors{Elle Weeks\affil{1}, Eli Tziperman\affil{1,2}}

\affiliation{1}{School of Engineering and Applied Sciences, Harvard University, Cambridge, Massachusetts}
\affiliation{2}{Department of Earth and Planetary Sciences, Harvard University, Cambridge, Massachusetts}

\correspondingauthor{Elle Weeks}{elleweeks@g.harvard.edu}


\begin{keypoints}
\item It is still not clear whether ENSO is a damped stochastically-driven oscillation or a self-sustained one.
\item Fitting the Recharge Oscillator model via multivariate linear regression systematically underestimates the growth rate of ENSO.
\item It is therefore challenging to diagnose whether the observed ENSO is in a damped or self-sustained regime using this approach.
\end{keypoints}

%
%


\begin{abstract}
The recharge oscillator (RO) model has been successfully used to understand different aspects of the El Niño-Southern Oscillation (ENSO). Fitting the RO to observations and climate model simulations consistently suggests that ENSO is a damped oscillator whose variability is sustained and made irregular by external weather noise. We investigate the methods that have been used to estimate the growth rate of ENSO by applying them to simulations of both damped and self-sustained RO regimes. We find that fitting a linear RO leads to parameters that imply a damped oscillator even when the fitted data were produced by a model that is self-sustained. Fitting a nonlinear RO also leads to a significant bias toward the damped regime. As such, it seems challenging to conclude whether ENSO is a damped or a self-sustained oscillation by fitting such models to observations, and the possibility that ENSO is self-sustained cannot be ruled out.
\end{abstract}

\section*{Plain Language Summary}
The El Niño Southern Oscillation (ENSO) is the largest driver of year-to-year global climate variability. An important question still unresolved is whether ENSO is driven by weather variability (noise, random forcing) or is self-sustained and would have existed without such random forcing. Addressing this often involves fitting a conceptual, often linear, model to observations of ENSO or to climate model output to estimate its characteristics such as the period and growth rate. Using this approach, previous studies typically estimate a negative growth rate suggesting that ENSO is a damped system requiring random wind events to sustain it. In this study, we investigate these methods for estimating the ENSO growth rate from observations and find that the growth rate is likely being substantially underestimated by such a fit. Our results imply that it is possible that the true ENSO growth rate could be positive implying that we cannot rule out that the ENSO cycle could exist without the need for external weather forcing.

%
%

\section{Introduction}

The El Niño Southern Oscillation (ENSO) is the dominant mode of interannual variability in the equatorial Pacific. ENSO teleconnections communicate its effects globally, making it a key driver of interannual climate variations \cite{timmermann2018nino}. General circulation models (GCMs) reproduce many key features of ENSO but are limited by several biases and are computationally expensive to run \cite{latif2001ensip, guilyardi2020enso, planton2021evaluating}. Additionally, future projections are inconsistent in their predictions of ENSO amplitude, period, and pattern across models \cite{vecchi2010nino, chen2017enso, maher2023future}. As such, simple conceptual models represent a useful tool for understanding ENSO. Several simple models have been developed and used to advance our understanding of the dynamics of the ENSO system \cite{neelin1998enso}, including the delayed oscillator \cite{suarez1988delayed, battisti1989interannual} and the recharge oscillator (RO) models \cite{jin1997Aequatorial, jin1997Bequatorial, vialard2024nino}. The recharge oscillator, which is the main tool used here, is represented by two ordinary differential equations describing the evolution of the ENSO sea surface temperature (SST) and equatorial heat content anomalies. The processes described by the model include the Bjerknes feedback in which SST anomalies intensify rapidly, the slow equatorial heat content adjustment mediated by Kelvin and Rossby waves, and the delayed oceanic feedback in which cooling occurs due to upwelling and westward currents \cite{jin1997Aequatorial}. Further work extended the RO to include stochastic forcing and nonlinear effects \cite{Schopf-Suarez-1988:vacillations, battisti1989interannual, jin1997Aequatorial, perez2005comparison, jin2007ensemble, vialard2024nino}.

Previous work has used the RO to study many aspects of ENSO dynamics, including the stability of the system \cite{burgers2005simplest, jin2006coupled, kim2011enso, kim2014enso}, the predictability of ENSO \cite{jin2007ensemble, frauen2012influences, levine2015annual}, the asymmetry between El Niño and La Niña events \cite{frauen2010enso}, and the diversity of ENSO events \cite{yu2016enso, wengel2018controls, capotondi2020enso}, among others \cite{jansen2009tropical, vijayeta2018evaluation}. One important application involves fitting the RO to observations or comprehensive climate model output to estimate the RO parameters, which can be used to quantify feedback strengths, the ENSO amplitude, or the stability \cite{burgers2005simplest, jansen2009tropical, frauen2012influences, wengel2018controls, vijayeta2018evaluation, vialard2024nino}. The stability of ENSO is determined by its growth rate---a negative growth rate implying a stable, damped regime and a positive growth rate implying an unstable, self-sustained regime. In particular, the Bjerknes‐Wyrtki‐Jin (BWJ) index allows one to quantify both the ENSO growth rate and period using the diagnosed RO parameters \cite{jin2006coupled, lu2018coupled, jin2020simple}. In practice, fitting the linear recharge oscillator to observations and climate model output consistently yields a negative growth rate, suggesting that ENSO is a damped oscillator sustained by stochastic forcing \cite{burgers2005simplest, wengel2018controls, vialard2024nino}. However, as carefully noted by \citeA{burgers2005simplest} and discussed in \citeA{jin1997Aequatorial}, weakly damped and slightly supercritical systems (i.e., positive growth rate near zero) may appear very similar in the presence of noise. Additionally, alternative models of ENSO suggest that the dynamics could instead be described by a self-sustained chaotic regime \cite{vallis1986nino, tziperman1994nino, tziperman1995irregularity, chang1996chaotic}.

In this work, we investigate the accuracy of methods that have been used to estimate the RO parameters and subsequently estimate the growth rate and period of ENSO from observations or climate model output. Using simulations of a nonlinear recharge oscillator in which the true RO parameters are known, we quantify the errors in the parameter estimates and determine if standard methods can distinguish between a damped regime and a self-sustained regime. We find that fitting a linear RO results in parameter estimates that imply a damped system, even when the fitted data were simulated by a self-sustained model. Applying these fitting methods over a range of parameter values shows that the growth rate is consistently underestimated, even when the model used to fit the data exactly matches the model used to generate them. Ultimately, these results point to the challenges in robustly differentiating between a damped or a self-sustained ENSO regime.

\section{Data and Methods}
\label{sec:data-and-methods}

The Recharge Oscillator model \cite{jin1997Aequatorial, burgers2005simplest, vialard2024nino} describes the evolution of the eastern equatorial Pacific SST and the equatorial Pacific heat content. In its simplest form, the RO is based on the following two linear ordinary differential equations, referred to as the linear recharge oscillator (LRO),
\begin{eqnarray}
    \label{eq:LRO_T}
    \frac{dT}{dt} &=& RT+F_1h+\zeta_T \\
    \label{eq:LRO_h}
    \frac{dh}{dt} &=& -\epsilon h-F_2T+\zeta_h,
\end{eqnarray}
where $T$ is the eastern equatorial Pacific SST anomaly, and $h$ is the equatorial Pacific thermocline depth anomaly. The parameter $R$ represents the Bjerknes feedback, $\epsilon$ captures the ocean basin adjustment mediated by Rossby and Kelvin waves, the $-F_2T$ term represents the effect of the Sverdrup transport on $h$, and the $F_1h$ term represents the delayed effects on $T$ through upwelling and westward currents. The stochastic forcing terms $\zeta_T$ and $\zeta_h$ primarily represent the weather forcings due to wind stress, including by westerly wind bursts that are known to depend on the SST \cite{Harrison-Vecchi-1997:westerly, jin1997Aequatorial, Tziperman-Yu-2007:quantifying, jin2020simple}.

The LRO describes a harmonic oscillator whose growth rate, $\gamma$, and frequency, $\omega$, are given by the real and complex parts of the Bjerknes–Wyrtki–Jin (BWJ) index, respectively \cite{jin2006coupled, lu2018coupled, jin2020simple}.
\begin{eqnarray}
    \label{eq:growth_rate}
    \gamma &=& \frac{R-\epsilon}{2} \\ 
    \label{eq:period}
    \omega &=& \sqrt{F_1F_2 - \frac{(R+\epsilon)^2}{4}}
\end{eqnarray}
Inferring a negative growth rate using the BWJ index implies a damped system, while inferring a positive growth rate implies a self-sustained one.

To better capture some of the complexities of the ENSO behavior such as the asymmetry between the El Ni\~no and La Ni\~na states, nonlinearities can be introduced into the RO model. We will refer to the model described by the following set of differential equations as the nonlinear recharge oscillator (NRO),
\begin{eqnarray}
    \label{eq:NRO_T}
    \frac{dT}{dt} &=& RT+F_1h+bT^2+cT^3+\zeta_T \\ 
    \label{eq:NRO_h}
    \frac{dh}{dt} &=& -\epsilon h-F_2T+\zeta_h.
\end{eqnarray}
Here, the quadratic nonlinearity $(bT^2,\,b>0)$ represents the physical processes that favor the growth of El Ni\~no relative to La Ni\~na leading to larger amplitude and shorter El Ni\~no events \cite{frauen2010enso}. The cubic nonlinearity $(cT^3,\,c<0)$ represents saturation effects on the ENSO amplitude \cite{Schopf-Suarez-1988:vacillations, battisti1989interannual, jin1997Aequatorial}.

In this work, we will explore the parameter values obtained by fitting both the LRO and the NRO equations to observations and simulated data.

For observations, we use the ORAS5 ocean reanalysis product from 1958 to 2020 \cite{copernicus2021oras5}. We use the monthly mean time series for the sea surface temperature (SST) averaged over the Ni\~no3 region (5\textdegree N to 5\textdegree S, 150\textdegree W to 90\textdegree W) and the thermocline depth defined as the 20\textdegree C isotherm depth averaged over the equatorial Pacific (5\textdegree N to 5\textdegree S, 120\textdegree E to 80\textdegree W). We compute monthly anomalies by subtracting the corresponding mean for each month over the full time series.

We generate time series of $T$ and $h$ for which we know the true model and parameter values by integrating the RO equations. We perform this integration using a forward Euler method with a time step of 24 hours and red noise stochastic forcing terms with a decorrelation timescale of 3 days following \citeA{vijayeta2018evaluation, wengel2018controls}. The integrations are performed using the NRO using equations (\ref{eq:NRO_T}, \ref{eq:NRO_h}). We simulate time series of daily anomalies and compute monthly averages of the simulated daily data to obtain monthly mean anomalies.

For the analysis shown below, we perform simulations for two cases, a damped oscillatory regime and a self-sustained oscillatory regime. The RO parameters for the damped parameter regime were calculated by fitting the NRO equations to monthly mean anomalies computed from the ORAS5 reanalysis product. They are consistent with previous work fitting the RO parameters from observations \cite{burgers2005simplest, wengel2018controls, vijayeta2018evaluation, vialard2024nino}. This damped oscillator has a decay timescale of 1.7 years and period of 3.7 years. For the self-sustained oscillator regime, we chose RO parameters such that the oscillator had a growth timescale of 11 years and shared features with the damped oscillator, such as the period, spectrum, and shape of the distributions of $T$ and $h$. The magnitude of the stochastic forcing terms, $\zeta_T$ and $\zeta_h$, are the same across all simulations. The RO parameters used in the simulations for these two cases are listed in Table~\ref{table:params}. For both regimes, we simulated 100 time series for a duration of 150 years each and perform our analyses using the last 100 years.

In addition, we simulate another set of 10,000 RO time series in which the values of all NRO parameters ($R$, $\epsilon$, $F_1$, $F_2$, $b$, $c$, std$(\zeta_T)$, and std$(\zeta_H)$) are varied randomly. Again, each time series was simulated for a duration of 150 years, and results are from the last 100 years.

\begin{table}
\caption{Parameter values used for simulating the damped and self-sustained RO.}
\label{table:params}
\centering
\begin{tabular}{l c c}
\hline
 \textbf{Parameters}  & \textbf{Damped} & \textbf{Self-sustained}  \\
\hline
  $R$ (1/year)  & -1.08   & 0.45 \\
  $\epsilon$ (1/year) & 0.13  & 0.26 \\
  $F_1$ (K/m/year) & 0.19  & 0.21 \\
  $F_2$ (m/K/year) & 16.7  & 15.0 \\
  $b$ (1/K/year) & 0.28 & 0.79 \\
  $c$ (1/K$^2$/year) & -0.028 & -1.35 \\
  std$(\zeta_T)$ (K/year)  & 3.11 & 2.63 \\
  std$(\zeta_h)$ (m/year) & 20.1  & 13.8 \\
\hline
\end{tabular}
\end{table}

Finally, we estimate the RO parameters from observations and simulated time series. We follow previous studies and estimate the RO parameters by fitting the LRO equations (\ref{eq:LRO_T}, \ref{eq:LRO_h}) via multivariate linear regression of monthly mean tendencies of $T$ and $h$ against monthly mean anomalies of $T$ and $h$ from observations or model output \cite{burgers2005simplest, vijayeta2018evaluation, wengel2018controls, vialard2024nino}. Following the above previous studies that fitted the RO to observations and model output, we compute the monthly mean tendencies using a forward difference approximation, e.g., $(T_{n+1}-T_n)/\Delta t = RT_n+F_1h_n$, where $n$ indicates monthly values and $\Delta t$ is one month. We perform linear regression of the monthly mean tendencies of $T$ on monthly mean anomalies of $T$ and $h$ to obtain estimates for parameters $R$ and $F_1$. We perform a separate linear regression of the monthly mean tendencies of $h$ on monthly mean anomalies of $h$ and $T$ to obtain estimates for parameters $\epsilon$ and $F_2$. The standard deviation of the stochastic forcing terms can then be approximated by the standard deviation of the residuals of the linear regression fit \cite{vijayeta2018evaluation, wengel2018controls, vialard2024nino}.

We similarly fit the NRO parameters using Equations~(\ref{eq:NRO_T} and \ref{eq:NRO_h}) as in \citeA{vialard2024nino}. We perform linear regression of the monthly mean tendencies of $T$ on $T$, $h$, $T^2$, $T^3$ to obtain estimates for the parameters $R$, $F_1$, $b$, and $c$. Estimates for $\epsilon$ and $F_2$ are obtained using the same fit described above, since (\ref{eq:LRO_h}) and (\ref{eq:NRO_h}) are identical.

Given estimates for $R$, $\epsilon$, $F_1$, and $F_2$ obtained by fitting either the LRO or the NRO, we estimate the growth rate and period using the BWJ index \cite<Equations~\ref{eq:growth_rate} and \ref{eq:period},>[]{jin2006coupled, lu2018coupled, jin2020simple}.

\section{Results}
\label{sec:results}

We now describe the results regarding the accuracy of estimating the RO parameters from simulated time series, where we know the true model parameter values used to generate the time series. We then consider the implications for our ability to distinguish between a damped and a self-sustained oscillation in the presence of noise as pertains to ENSO in the climate system.

\begin{figure}[!tbhp]
    \centering
    \includegraphics[width=\textwidth]{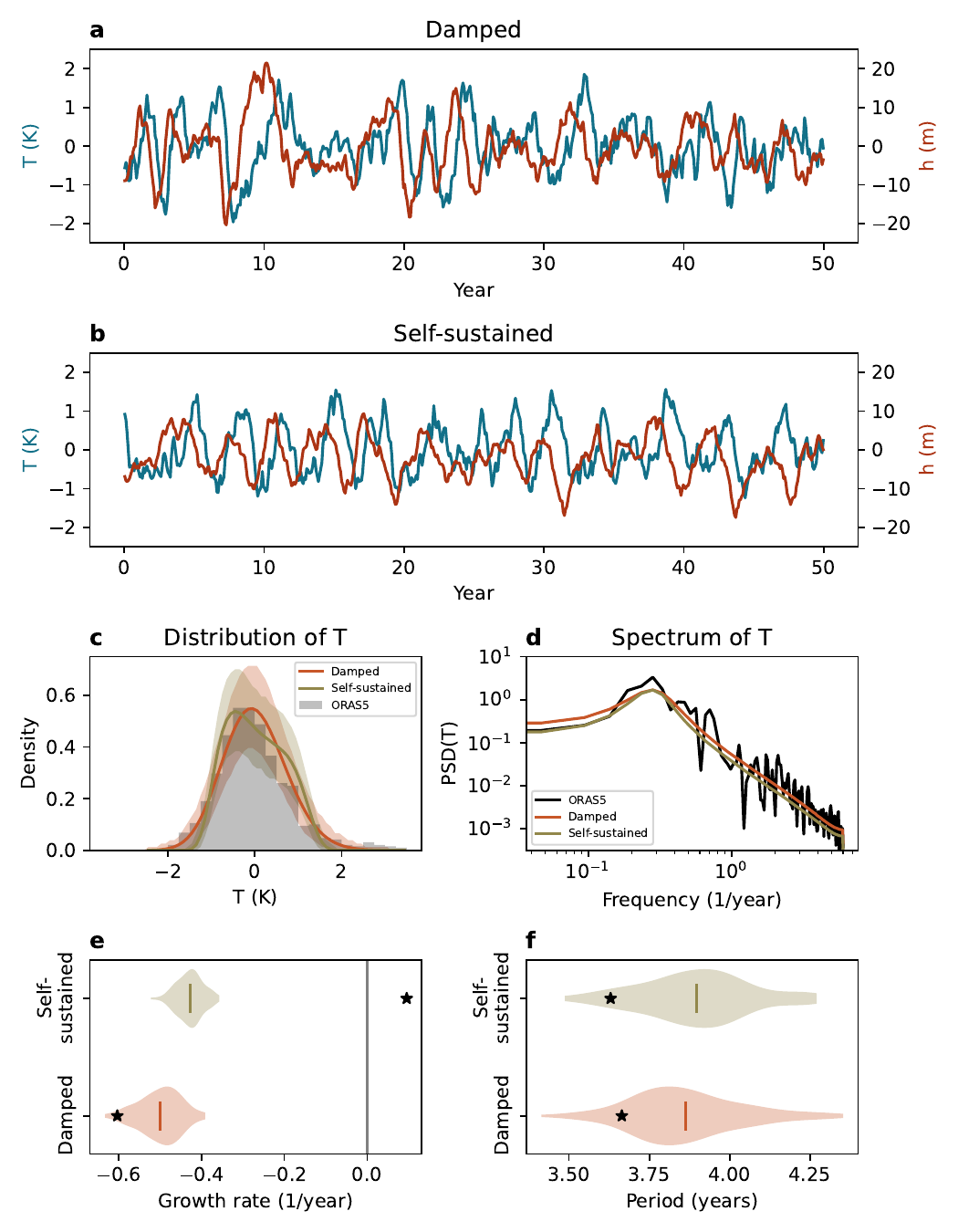}
    \caption{Comparison between a simulated damped recharge oscillator, a simulated self-sustained recharge oscillator, and observations. (a) Time series of $T$ (blue) and $h$ (red) for a damped system. (b) Time series of $T$ (blue) and $h$ (red) for a self-sustained system. Comparison of the distributions (c) and spectra (d) of $T$. Estimates of the growth rate (e) and period (f) obtained by fitting the LRO to the simulated time series. The stars indicate the true parameter values used to produce the fitted time series.}
    \label{fig:damp_vs_sustained}
\end{figure}

We first reiterate the known result \cite{jin1997Aequatorial, burgers2005simplest} that both a damped and a self-sustained RO can match the characteristics of ENSO found in observations (Figure~\ref{fig:damp_vs_sustained}). We analyze simulated time series of the mean monthly anomalies, $T$ and $h$, for a damped system with a decay timescale of 1.7 years (Figure~\ref{fig:damp_vs_sustained}a) and for a self-sustained system with a growth timescale of 11 years (Figure~\ref{fig:damp_vs_sustained}b). We compare the distributions of the simulated monthly values for $T$ to those from observations and find similar centers and spread among the three distributions (Figure~\ref{fig:damp_vs_sustained}c). Similarly, we compute the spectrum of the simulated and observed time series for $T$ and find strong agreement (Figure~\ref{fig:damp_vs_sustained}d). All three spectra show a peak corresponding to an average period of around 3.7 years, consistent with an ENSO period of approximately 2--5 years \cite{jiang2021nino}.

Next, we estimate the RO parameters $R$, $\epsilon$, $F_1$, and $F_2$ by fitting the LRO equations (\ref{eq:LRO_T}, \ref{eq:LRO_h}) to our simulated time series for $T$ and $h$ and compute the estimated growth rate and period (\ref{eq:growth_rate}, \ref{eq:period}). The distributions of the estimated growth rate and period across 100 simulations for both the damped and self-sustained systems are shown in Figure~\ref{fig:damp_vs_sustained}e,f. We find that, for the damped system, the fit tends to estimate weaker damping than the true value and, for the self-sustained system (positive growth rate), the fit consistently estimates a damped system (negative estimated growth rate) across all simulations (Figure~\ref{fig:damp_vs_sustained}e). The estimates of the period are more accurate, with most estimates falling within 6 months of the true simulated period (Figure~\ref{fig:damp_vs_sustained}f). These results suggest that fitting the linear RO to data may not be able to accurately estimate the growth rate and may estimate a damped system (negative growth rate) when the system is self-sustained (positive growth rate). 

\begin{figure}[!tbh]
    \centering
    \includegraphics[width=.85\textwidth]{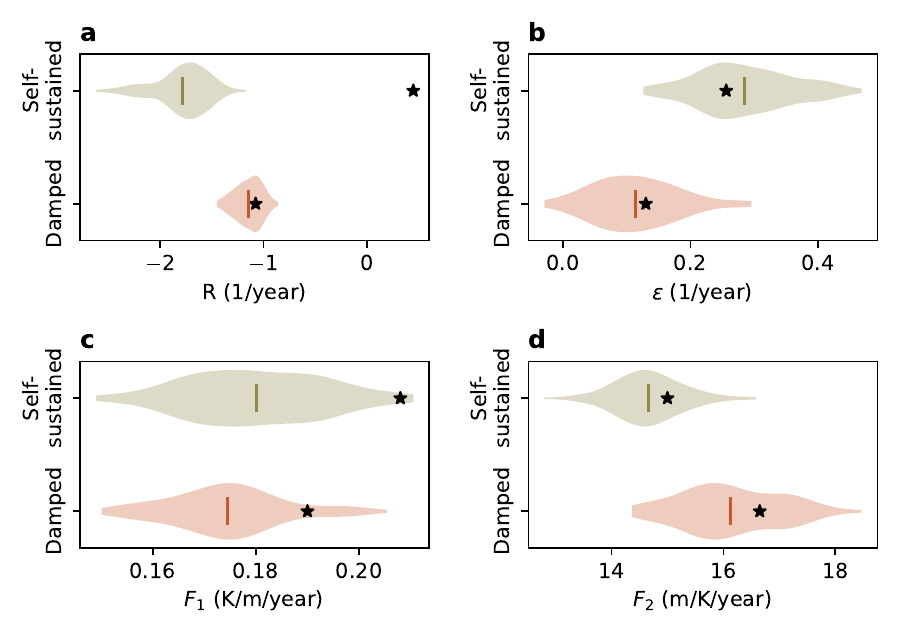}
    \caption{Estimates of the RO parameters---(a) R, (b) $\epsilon$, (c) $F_1$, (d) $F_2$---obtained by fitting the LRO to the simulated time series. Stars indicate the true parameter values.}
    \label{fig:param_est}
\end{figure}

To see which LRO parameters are not well estimated by the fit, Figure~\ref{fig:param_est} shows distributions of the estimates for these four model parameter values across 100 simulated time series compared to their true value. For both the damped and self-sustained cases, the true value of $\epsilon$, $F_1$, and $F_2$ fall within the range of estimated values (Figure~\ref{fig:param_est}b,c,d). However, for all parameters, we note that the mean estimate over many simulations does not appear to match the true value. The estimates for $R$ show large errors, particularly for the self-sustained system, where the estimates of $R$ have the incorrect sign for all simulations (Figure~\ref{fig:param_est}a). The growth rate depends on $R$ and $\epsilon$ (\ref{eq:growth_rate}), so these large errors in the estimates for $R$ translate into the large errors in the estimate of the growth rate (Figure~\ref{fig:damp_vs_sustained}e). In particular, for the self-sustained case, the large negative estimated value for $R$ when the true value is positive results in the estimated growth rate suggesting a damped system despite the simulated system being self-sustained. The period depends on all four parameters, with the product $F_1F_2$ being the dominant term under the radical in (\ref{eq:period}). Therefore, the period is less sensitive to the errors in estimates for $R$, and estimates of the period tend to be closer to the true value due to the smaller errors in the estimates of $F_1$ and $F_2$.

Fitting the LRO using a longer time series reduces the variance in the estimates, but the error in the mean of the estimates persists (Figure~S1). These results suggest that fitting an LRO to the time series may lead to a biased estimate for the parameters of interest, i.e. the mean of the parameter estimates does not approach the true value even when more simulations are performed or when longer time series are used. The bias in the estimator may have several possible explanations, including mismatch between the models used to produce and fit the data (which is an expected difficulty when fitting the LRO to GCMs or observations), averaging over nonlinearities in the model, and autocorrelation in the noise term \cite<Section 3.3.3,>{james2013introduction}. We now discuss how addressing each of these possible sources of error affects our estimates.

\begin{figure}[!tbh]
    \centering
    \includegraphics[width=.85\textwidth]{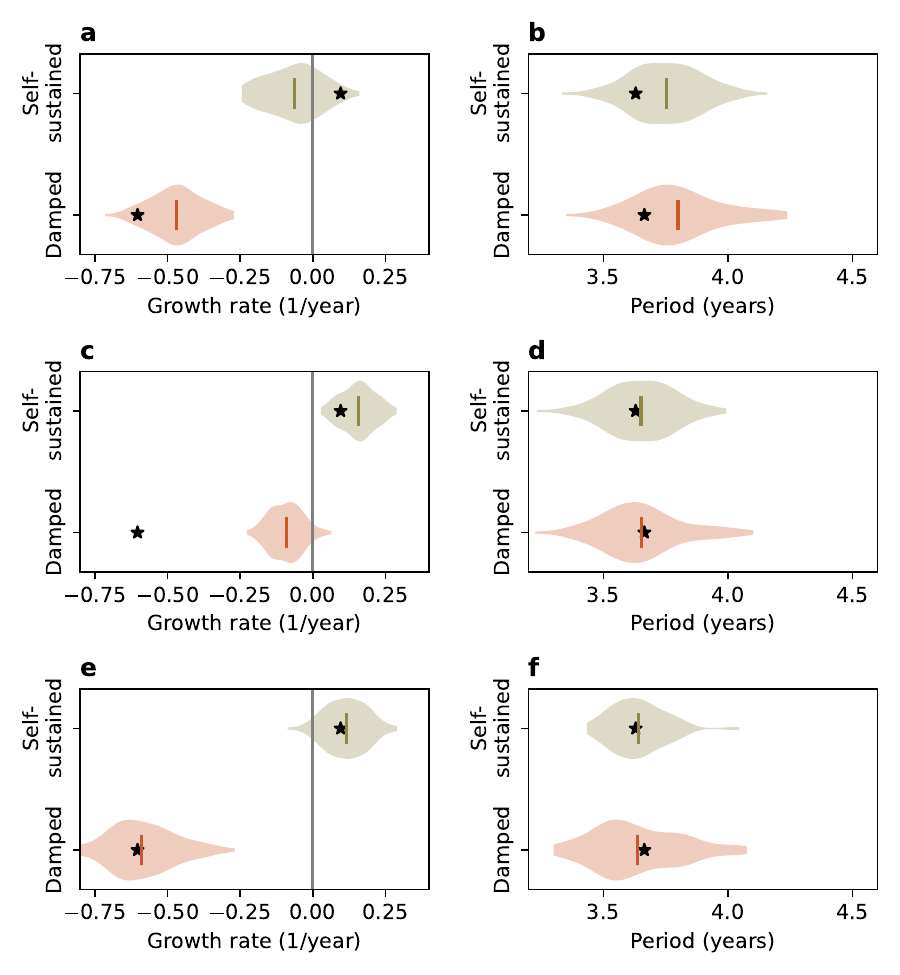}
    \caption{Estimates of the growth rate (left) and period (right) obtained by variations of fitting the RO equations to the simulated time series. Stars indicate the true parameter values. (a,b) Parameters estimated by fitting the NRO equations to monthly means. (c,d) As in (a,b) but NRO equations fit to daily means. (e,f) As in (c,d) but time series simulated with white noise.}
    \label{fig:bias_source}
\end{figure}

Figure~\ref{fig:bias_source} analyzes the source of the bias in the estimated model parameters. First, the model we fit was chosen intentionally at this first step not to be identical to the model that we use to generate the data. We simulate the time series for $T$ and $h$ according to the NRO equations to create a time series that recreates the asymmetries seen in observations of ENSO, but we fit the LRO equations following previous studies \cite{burgers2005simplest, vijayeta2018evaluation, wengel2018controls, vialard2024nino}. Estimating the RO parameters by fitting the same NRO equations that were used to generate the data improves the estimates of the growth rate and the period, but notable errors remain (Figure~\ref{fig:bias_source}a,b). The true values for the growth rate and period fall within the range of predicted values, but the means of our estimates do not appear to match the true value. In particular, the mean estimate of the growth rate for the self-sustained case is still negative, incorrectly suggesting a damped system.

Second, the model used to generate the data for the analysis (\ref{eq:NRO_T}, \ref{eq:NRO_h}) includes two nonlinear terms ($bT^2$ and $cT^3$). When fitting the model using monthly mean data, we follow the procedure used by previous studies and compute the monthly mean anomalies $T$ and $h$ then apply the nonlinearity instead of computing the nonlinear term and then computing the monthly mean, which may introduce errors. Using more frequent data to perform the fit and compute parameter estimates should reduce the errors due to averaging over the nonlinearity. Fitting the NRO equations using daily data instead of monthly again improves our estimates of the growth rate but does not eliminate the bias entirely (Figure~\ref{fig:bias_source}c).

Finally, ordinary least squares is only guaranteed to be an unbiased estimator when the errors are homoscedastic (i.e., the variance of the noise is constant) and serially uncorrelated \cite<Section 3.3.3,>{james2013introduction}. Yet, the stochastic forcing we simulated has a decorrelation time of three days, violating this condition even when we use daily data. We test how violating this condition affects our estimates by simulating daily time series for $T$ and $h$ using stochastic forcing ($\zeta_T$ and $\zeta_h$) that is uncorrelated in time. We then fit the NRO equations to the daily data. Only under these conditions, including no model mismatch, high temporal resolution data, and uncorrelated noise, does this method of fitting the RO parameters return unbiased estimates (Figure~\ref{fig:bias_source}e,f).

We cannot expect to satisfactorily address all of these sources of error when fitting the RO to observations. First, we do not know the exact dynamics for ENSO in the climate system. The RO is a very useful conceptual model but is not meant to capture the full complexity of ENSO, so we can always expect some mismatch between the model and the observations. Additionally, it is not possible to remove the correlation in the noise of the real system, which is representative of weather noise and westerly wind bursts, whose correlation time is of the order of a week. Given these constraints, we expect it would be difficult to correctly diagnose whether the observed ENSO is damped or self-sustained using the analysis tools examined here.

\begin{figure}[!tbh]
    \centering
    \includegraphics[width=\textwidth]{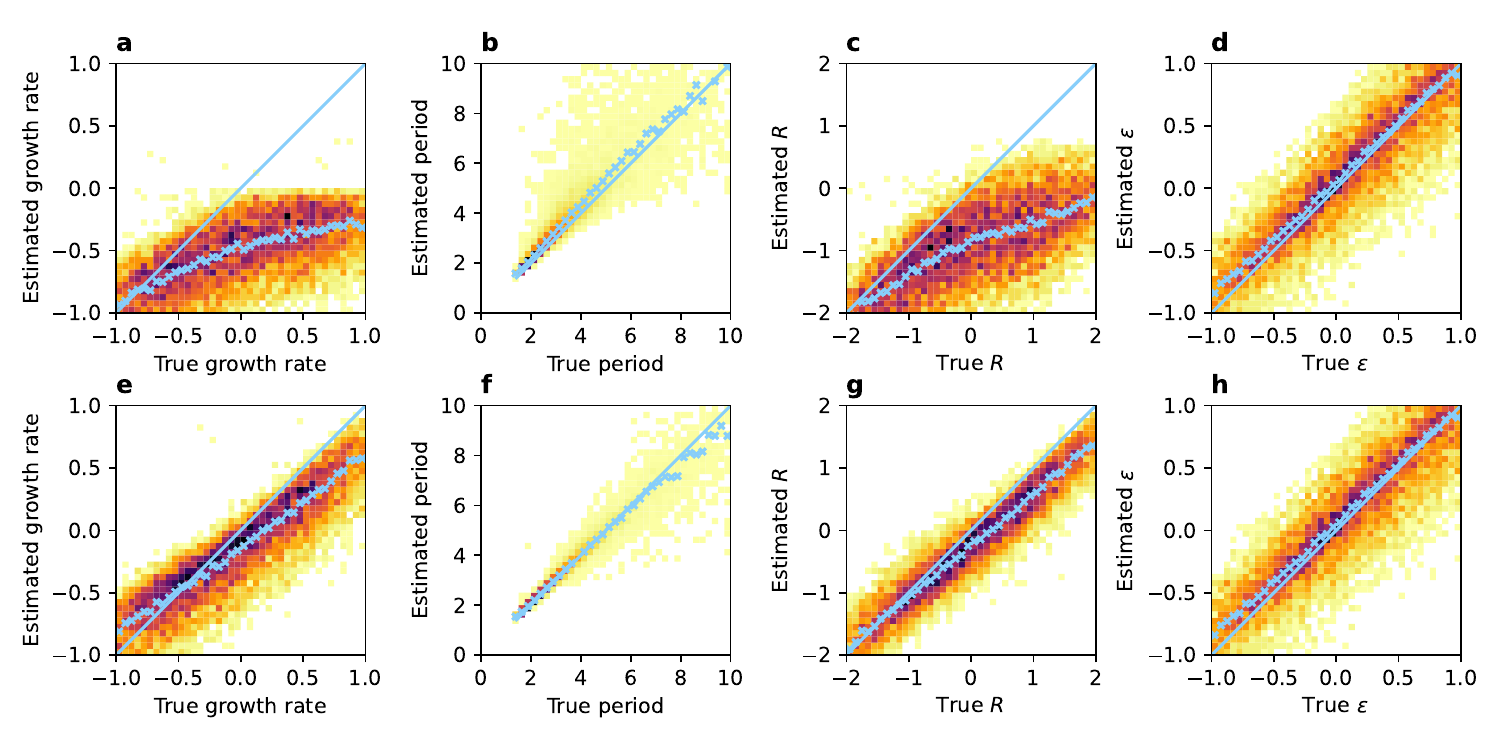}
    \caption{Comparison of estimates of the RO parameters---(a,e) growth rate, (b,f) period, (c,g) $R$, (d,h) $\epsilon$---obtained by fitting the LRO (a-d) or NRO (e-h) to the simulated time series with the true values. Time series were simulated according to the NRO varying the true values of $R$, $\epsilon$, $F_1$, and $F_2$. The one-to-one line is shown in blue. $\times$s indicate binned mean estimates.}
    \label{fig:param_sweep}
\end{figure}

Up to this point, we have considered two sets of representative parameter values corresponding to a self-sustained and a damped regime. We do not know the RO parameter values that would most accurately describe the observed ENSO. We now consider a broad range of RO parameter values for $R$, $\epsilon$, $F_1$, $F_2$, $b$, $c$, std$(\zeta_T)$, and std$(\zeta_H)$, producing corresponding time series for each set of parameters. We simulated 10,000 such time series for $T$ and $h$ in which the model period ranges from 0 to 10 years, and the growth rate ranges from $-1$ to 1 year$^{-1}$. We then followed the same method as above for estimating $R$, $\epsilon$, $F_1$, and $F_2$ by fitting the LRO equations to the monthly mean anomalies. Figure~\ref{fig:param_sweep}a--d shows the comparison between the true and estimated parameter values, growth rate, and period. We find that this fit of the RO parameters underestimates the true growth rate and overestimates the true period on average (Figure~\ref{fig:param_sweep}a,b). Most strikingly, when the true growth rate is positive, the estimated growth rate is effectively always negative (Figure~\ref{fig:param_sweep}a). This means that when the true system is self-sustained, this approach will lead to the incorrect conclusion that the system is damped. The error in the growth rate appears to mainly be due to errors in the estimated value for $R$ (Figure~\ref{fig:param_sweep}c). When the true growth rates is greater than zero, the estimated growth rate is consistently around $-0.27$ year$^{-1}$ (a decay scale of 4 years). As a reminder, when the LRO is fit to observations, the estimated growth rate is $-0.53$ year$^{-1}$. In short, it appears that estimating a negative growth rate by fitting the LRO may incorrectly suggest a damped system, even when the true dynamics are self-sustained.

Having previously shown that fitting the LRO when the model generating the data was nonlinear contributes to the errors we find in estimating the RO parameters, we now repeat our previous analysis but fit the NRO equations to our simulated time series. This analysis, allows us to explore the accuracy of our parameter estimates when the model generating the data exactly matches the model that we fit. While the errors are smaller now, this fit still tends to underestimate the value of $R$ (Figure~\ref{fig:param_sweep}g) and overestimate the value of $\epsilon$ (Figure~\ref{fig:param_sweep}h), both of which contribute to underestimating the growth rate (Figure~\ref{fig:param_sweep}e). Only once the true growth rate is above 0.23 year$^{-1}$ do we estimate a positive growth rate on average. When we fit the NRO to observations, we estimate a growth rate of $-0.6$ year$^{-1}$. 

When fitting observations, one expects the dynamics to have some differences from the fitted model. We test the sensitivity of the accuracy of the parameter estimates to this by fitting a different nonlinearity than the nonlinearity included in the model generating the data. For this test, we remove the $cT^3$ term from the equation for the evolution of the SST and simulate the time series with a $ch^3$ term in the equation for the evolution of the thermocline depth instead. Then, we fit the nonlinear model with the $cT^3$ term in the equation for the evolution of the SST as described by (\ref{eq:NRO_T}, \ref{eq:NRO_h}). We find that fitting the wrong nonlinearity (as is likely to be the case when fitting a simple model to observations or climate model output) can yield errors in estimates of the growth rate as large as those found by fitting the LRO to simulations produced using the NRO (Figure~S2).

These results suggest that the parameter errors found by fitting the correct model represent a lower bound for the errors. The errors when fitting to observations are expected to be even larger, because we cannot expect the RO model to perfectly represent the observed dynamics. Therefore, fitting the RO to observations or climate model output likely underestimates the true ENSO growth rate.

\section{Conclusions}
\label{sec:conclusions}

We have investigated the potential for characterizing ENSO as a damped or a self-sustained regime by fitting the recharge oscillator (RO) to observations or climate model output via multivariate linear regression. Previous work fitting the RO, both linear and nonlinear, to observations and climate model output consistently yields results that suggested ENSO is a damped oscillator. Here, we explore the robustness of this approach. By leveraging simulations of RO time series for which we know the true parameter values, we attempt to quantify the error in the parameter estimates obtained by fitting the RO equations. We then considered how these errors in the RO parameters affect the inferred growth rate and period of ENSO.

We have found that estimating the ENSO growth rate using this fitting approach consistently underestimates the growth rate of ENSO. Not surprisingly, the errors in the estimates are greater when the fitted model is not identical to the model generating the data, as would be the case when fitting observations and climate models. We found that a damped ENSO is often identified even when the simulated system was self-sustained. Overall, our results suggest that it would be challenging to robustly differentiate between damped and self-sustained ENSO regimes by estimating the growth rate from the fitted RO parameters. This problem is especially severe when a linear RO is fitted, where the estimate is effectively never self-sustained over a wide range of parameters explored for the model used to generate the fitted data.

A less damped ENSO regime suggested buy our results is consistent with estimates of the ENSO growth rate in \citeA{jin2020simple} based on analytic formulae for the RO parameters. Their estimates yielded a growth rate of $-0.3$ year$^{-1}$ (a 3-year decay time scale), which they note implied that the ENSO dynamics are near critical. This estimate represents much weaker damping than typical estimates from fitting the RO parameters via linear regression. For example, we estimated the ENSO growth rate to be $-0.53$ year$^{-1}$ when fitting the LRO to observations, (1.9 year damping time scale, consistent with \citeA{burgers2005simplest, wengel2018controls, vijayeta2018evaluation, vialard2024nino}). \citeA{jin2020simple} attribute the differences to nonlinearities being captured in the linear fitting method but not in their analytic formulae. Our results suggest that the difference could also be due to a biased estimation of the growth rate when fitting the RO to observations.

This study is based on input produced by RO equations to test the fit by a RO, and therefore provides a lower bound for the errors in the parameter estimates. It would be more challenging to estimate the expected parameter errors one would find by fitting the RO to observations or climate model output, as the extent of model mismatch between the RO and observations is unknown. We followed previous studies and fitted both linear and nonlinear RO models, yet other variations on the NRO exist, such as models that include a multiplicative noise term or allow the value of $F_1$ to depend on the sign of $h$. Fitting other variants of the RO may yield different estimates for the error lower bound.

In summary, our results show that the ENSO growth rate is likely underestimated by fitting the RO to observations or climate model output. This underestimation can yield a negative estimated growth rate (damped oscillatory regime) even when the dynamics are characterized by a positive growth rate (self-sustained regime). The challenges in distinguishing between a damped and a slightly supercritical system were clearly noted by \citeA{burgers2005simplest}. Similarly, it is well-known that a self-sustained chaotic description of ENSO is difficult to distinguish from a damped oscillatory stochastically driven regime, although with a long enough time series there are tools that allow one to distinguish between the two \cite{tziperman1994nino}. Our results further highlight the potential challenges of a specific method that has been used to estimate ENSO's growth rate and, ultimately, show that it may therefore not be straightforward to conclude whether ENSO is a damped or self-sustained oscillator.

\clearpage



%
%

\section*{Open Research Section}
The ORAS5 reanalysis data used for this paper is available at https://cds.climate.copernicus.eu/datasets/reanalysis-oras5?tab=download, DOI: 10.24381/cds.67e8eeb7. Numerical simulations and python code used to generate the results and figures in this work can be found at Zenodo at https://doi.org/10.5281/zenodo.15121763.

\acknowledgments
This work was funded by Department of Energy (DOE) Office of Science Biological and Environmental Research grant DE-SC0023134. ET thanks the Weizmann Institute of Science for its hospitality during parts of this work.

%

\bibliography{references}

\end{document}